%% file: Draft.tex
\documentclass[11pt,english,longbibliography]{article}
\usepackage{jheppub}
\input{packages.tex}
\usepackage{indentfirst}
\usepackage{yfonts}
\usepackage{pgfplots}
\pgfplotsset{compat=1.18}
\usepgfplotslibrary{fillbetween}
\usetikzlibrary{patterns, positioning, fadings, arrows.meta, calc,decorations.text}
\title{\fontsize{50}{30}\selectfont\textfrak{ Regge's Inferno}}

\author[a]{Zohar Komargodski}
\author[b]{Alessio Miscioscia}
\author[a]{Fedor K. Popov}

\affiliation[a]{Simons Center for Geometry and Physics, Stony Brook University, Stony Brook, NY}
\affiliation[b]{C. N. Yang Institute for Theoretical Physics, Stony Brook University, Stony Brook, NY 11794,USA}

\abstract{
We study large-spin operators in conformal field theories (CFTs) in spacetime dimensions $d>2$ by placing the theory on appropriate pp-wave backgrounds. We show that these geometries admit Heisenberg-group symmetries, and that these symmetries, combined with locality of quantum fields on such spacetimes, impose strong constraints on the asymptotic spectrum in the large-spin limit. The pp-wave backgrounds probe both the small-twist regime, corresponding to the Regge or light-cone bootstrap, and a strongly coupled regime of large twist. Finally, we demonstrate that causality  (or the requirement that the energy be bounded from below) leads to a new unitarity bound in $3+1$ dimensions.

}

\begin{document} 

 \vspace*{-.6in}
 \begin{flushright}
     YITP-SB-2026-05
 \end{flushright}

\maketitle
\section{Introduction and Summary}

The central objects of any quantum field theory are local operators. These
local operators in a conformal field theory (CFT) are characterized by their scaling dimensions, Lorentz spins, and charges under global symmetries,
\begin{equation}{\cal O}_{\Delta, J_i,Q_a}~
\end{equation}
A major goal in the study of CFTs is to determine the spectrum of the operators, i.e. the spectrum of critical exponent $\Delta$. This is a highly non-trivial problem, solvable in full detail only in very special cases. 

Various aspects of this problem are well understood. Low-lying operators are carved out by the numerical bootstrap techniques~\cite{Poland:2018epd}. Large charge operators with the smallest possible $\Delta$ are often organized into a superfluid effective field theory~\cite{Hellerman:2015nra,Monin:2016jmo,Jafferis:2017zna,Badel:2019oxl,Grassi:2019txd, Cuomo:2021cnb,Cuomo:2022kio,Choi:2025tql}. Large spin operators furnish a weakly-coupled Fock space at low twist and they can be controlled by the analytic bootstrap~\cite{Alday:2007mf,Komargodski:2012ek,Fitzpatrick:2012yx,Fitzpatrick:2014vua, Simmons-Duffin:2016gjk,Caron-Huot:2017vep,Fardelli:2025eun,Pal:2022vqc,Kaviraj:2022wbw,vanRees:2024xkb,Harris:2024nmr,Fardelli:2025fkn}, see also the review~\cite{Bissi:2022mrs}. The main purpose of this paper is to reconsider the operators of large spin.

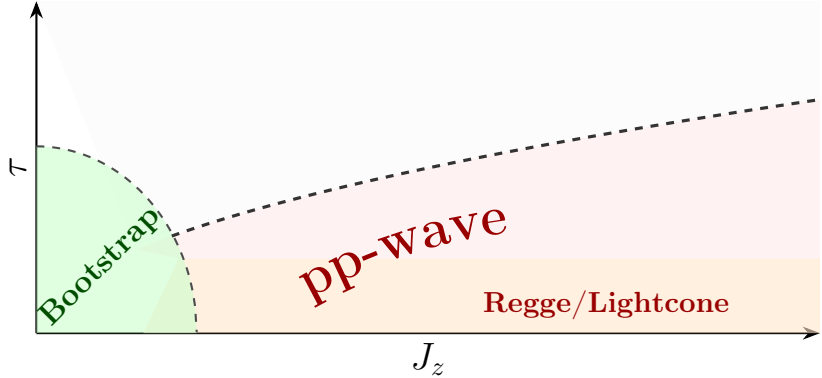
\begin{figure}
    \centering
    \begin{tikzpicture}
        \begin{axis}[
            xlabel={\Large $J_z$},
            ylabel={\Large $\tau$},
            xmin=0, xmax=22,
            ymin=0, ymax=8,
            axis lines=left,
            axis line style={-Stealth, thick},
            ticks=none,
            width=12cm,
            height=6cm,
            enlargelimits=false,
            clip=false
        ]

            \addplot[name path=curve, domain=2.8:22, samples=100, draw=none] {1.2*sqrt(x)};
            
            \path[name path=reggelimit] (axis cs:4,1.8) -- (axis cs:22,1.8);

            \path[name path=bottom] (axis cs:3,0) -- (axis cs:22,0);

            \addplot[orange!20, opacity=0.6] fill between[of=bottom and reggelimit];

            \addplot[red!10, opacity=0.5] fill between[of=reggelimit and curve];

            \path[name path=top] (axis cs:0,8) -- (axis cs:22,8);
            
            \addplot[gray!5, opacity=0.5] fill between[of=curve and top];
            \addplot[
                draw=none, 
                fill=green!20, 
                opacity=0.6, 
                domain=0:4.5, 
                samples=100
            ] 
            {sqrt(20.25 - x^2)} \closedcycle;
            
            \node[rotate=45, font=\bfseries\large, color=green!30!black, anchor=center] 
                at (axis cs:1.8, 1.6) {Bootstrap};
            \addplot[
                dashed, 
                thick, 
                color=black!70,
                domain=0:4.5, 
                samples=100
            ] 
            {sqrt(20.25 - x^2)};

            \addplot[
                domain=3.8:22,
                samples=100,
                dashed,
                very thick,
                color=black!80
            ]
            {1.2*sqrt(x)};

            \path[
                decoration={
                    text along path,
                    text={|\bfseries\color{red!60!black}|Regge/Lightcone},
                    text align=center
                },
                decorate
            ] 
            (axis cs:12, 0.5) to[bend left=0] (axis cs:20, 0.5);    

           \path[
                decoration={
                    text along path,
                    text={|\huge\bfseries\color{red!60!black}|pp-wave},
                    text align=center
                },
                decorate
            ] 
            (axis cs:7, 0.5) to[bend left=12] (axis cs:14, 2.5);

        \end{axis}
    \end{tikzpicture}
    \caption{A cartoon of different regimes in $(2+1)$-dimensional CFTs in the $(\tau, J_z)$ plane. 
Numerical bootstrap techniques are effective in probing the low-twist, low-spin regime, while analytic bootstrap methods (the Lightcone bootstrap) capture the large-spin sector at small twist. 
The pp wave geometry  developed in this work captures operators 
with large spin and arbitrarily large twist with  $\tau/\sqrt{J_z}$ fixed at large $J_z$.}
    \label{fig:phase}
\end{figure}

Our notation would be that in 2+1 dimensions operators carry the angular momentum $J_z$ corresponding to rotations around the $z$ axis. In 3+1 dimensions the operators are labeled by the angular momenta $J_1,J_2$ corresponding to rotations around the planes $12$ and $34$, respectively.  The twist in 2+1 dimensions is defined by $\tau=\Delta-J_z$ while in 3+1 dimensions it is defined by $\tau=\Delta-J_1-J_2$. 

At large spin we have the following simplification. Consider any primary operator ${\cal O}$, and consider multi-twist operators of the form 
\begin{alignat}{2}
\label{FOCKstatesintro}2+1 \ d: &\quad & \partial_+^{m_1}{\cal O}\cdots \partial_+^{m_n}{\cal O}, \\
\label{FOCKstatesintro2}3+1 \ d: &\quad & \partial_+^{(1)p_1}\partial_+^{(2)q_1}{\cal O}\cdots \partial_+^{(1)p_n}\partial_+^{(2)q_n}{\cal O}~,
\end{alignat}
where $\partial_+$ raises $J_z$ by one and $\partial_+^{(1)},\partial_+^{(2)}$ raise $J_1,J_2$ by one, respectively.  
These operators behave as weakly coupled partons:  their scaling dimensions  simply add up
$\Delta=n\Delta_0+\sum_{i=1}^n m_i$, and $\Delta=n\Delta_0+\sum_{i=1}^n p_i+\sum_{i=1}^n q_i$, up to corrections which are small for $m_i,p_i,q_i\gg1$. Therefore, unlike the generic case in CFT where composite operators receive large corrections to their scaling dimensions, these high-spin operators behave as nearly-free partons.

As we increase the number of partons, $n$, the interactions  become stronger. Eventually, the Fock space ceases to be a good approximation to the eigenstates of the dilatation operator. To estimate when the  parton approximation ceases to be a good description we can use the   lightcone bootstrap to calculate corrections to the scaling dimensions of the composite operators above. In any number of dimensions $d+1$, we find that if only one angular momentum is large, 
the interactions among the partons become large when \begin{equation}
\label{strong1}\tau_{\rm strong}\sim \sqrt J_z~.
\end{equation} 
While in 3+1 dimensions, if both angular momenta are large (with a fixed ratio), the interactions become strong when\footnote{The breakdown of the Fock space picture could potentially happen earlier in some cases.} \begin{equation}\label{strong2}\tau_{\rm strong}\sim (J_1J_2)^{1/3}~.
\end{equation}
An interesting regime that includes the weakly-coupled parton states as well as some strongly interacting states with large spin is obtained by taking 
\begin{equation}\label{scalingone} 
J_z\to \infty~,\quad {\tau\over\sqrt J_z}={\rm fixed}~, 
\end{equation}
in any spacetime dimension, as long as only one angular momentum  is large, or 
\begin{equation}\label{scalingtwo} 
J_{1,2}\to \infty~,\quad {J_1\over J_2}={\rm fixed}~,\quad  {\tau\over(J_1J_2)^{1/3}}={\rm fixed}~, 
\end{equation}
in 3+1 dimensions if both angular momenta are large. These interesting scaling regimes appeared for the first time in~\cite{SemiUniversality}, which inspired our work.

These scaling regimes include the weakly coupled partons, which have small $\tau/\sqrt J_z$ and $\tau/(J_1J_2)^{1/3}$, and they also include strongly coupled fast spinning states which are not easily accessible within the light-cone bootstrap, see Figure~\ref{fig:phase}.

\begin{figure}
    \centering
    \begin{tikzpicture}[scale=2, line cap=round, line join=round]
  
  \draw[thick] (0,0) circle (1);

  \draw[thick, variable=\t, domain=190:350, samples=200]
    plot ({cos(\t)}, {0.3*sin(\t) + 0.18});
    
    \draw[<->, thick] (1.15,-0.30) -- (1.15,0.15);
    \node[right] at (1.15,-0) {$\delta\theta \sim J_z^{-\frac12}$};
  
  \draw[thick, variable=\t, domain=195:345, samples=200]
    plot ({cos(\t)}, {0.3*sin(\t) - 0.18});
\foreach \t in {240,255,270,285,300}{
  \fill ({cos(\t)}, {0.35*sin(\t) + 0.05}) circle (0.03);
}
\draw[thick, postaction={decorate},
  decoration={markings, mark=at position 1 with {\arrow{>}}}]
  plot[variable=\t, domain=310:340, samples=120]
  ({cos(\t)}, {0.35*sin(\t) + 0.05});

\draw[thick, postaction={decorate},
  decoration={markings, mark=at position 1 with {\arrow{>}}}]
  plot[variable=\t, domain=190:220, samples=120]
  ({cos(\t)}, {0.35*sin(\t) + 0.05});
\end{tikzpicture}
\caption{Highly boosted states correspond to wavefunctions that localize near the equator of $S^2$, within a narrow strip of angular width $\delta\theta\sim J_z^{-\frac12}$. }
    \label{fig:SphereS2}
\end{figure}
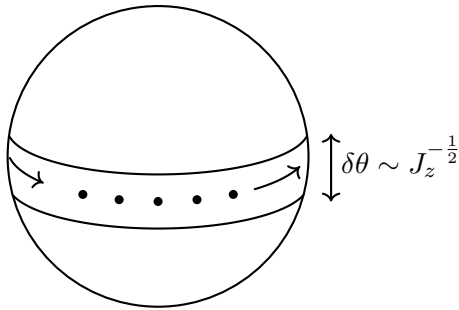

The main point of this paper is that the regimes~\eqref{scalingone},\eqref{scalingtwo} could be studied by placing the CFT on pp-wave geometries. In the case of a single large angular momentum in $d+1$ space-time dimensions we obtain the geometry 
\begin{gather}\label{onelargeintro}
    \boxed{ds^2 = -\left(1 + \sum^{d-1}_{i=1}x_i^2\right) dt^2 +  dx dt + \sum^{d-1}_{i=1} dx_i^2}~. 
\end{gather}
This pp-wave geometry is obtained by a null limit of the Lorentizan cylinder around the equator, see Figure~\ref{fig:SphereS2}. This metric is a Brinkmann pp-wave in \(2+1\) dimensions and was studied in various context, see \cite{Penrose1976,BlauEtAlPenrose2002,BMN2002,Metsaev2002,Blau:PlaneWavesPenroseLimits,Maldacena:2008wh}. 

The case of two large angular momenta in 3+1 dimensions~\eqref{scalingtwo} is not obtained by a scaling limit around a single null geodesic since the partons could be anywhere on the $S^3$. Nevertheless, one again obtains a pp-wave geometry  
\begin{gather} \label{eq:orig4dppintrod}
\boxed{ds^2 =-  dt^2  -  2 dt dv - 4 u dt dy + du^2 + dy^2}~
\end{gather}
We can get ~\eqref{eq:orig4dppintrod} by going to a rotating frame (at unit frequency) in the plane $(x_1,x_2)$ in~\eqref{onelargeintro}, or alternatively by doing a scaling limit to the family of null geodesics on $S^3$.
It is already interesting to note that going to a rotating frame in the plane $(x_1,x_2)$ does not produce regions with positive $g_{00}$, unlike in ordinary flat space. Indeed after this rotation the spacetime~\eqref{eq:orig4dppintrod} still has a global timelike Killing vector.

These space-times have a global timelike Killing vector,as well as interesting symmetries, which underlie the various large spin limits. 
The  metric~\eqref{onelargeintro} has an isometry group given by a Heisenberg group \(\mathbb H_{2(d-1)+1}\) together with time translations.
The metric~\eqref{eq:orig4dppintrod} has 
$\mathbb H_3 \rtimes SO(2)$ symmetry together with time translations. These symmetries lead to interesting constraints on the spectrum at the large spin limit as well as partition functions on these spaces. The Heisenberg groups above have nontrivial subgroups commuting with time translations. These subgroups survive at finite temperature and will be useful for us to constrain the thermodynamic properties of the systems. 

 In $d=4$, the presently known unitarity bounds do not a priori exclude operators with
$\tau = \Delta - J_1 - J_2 < 0$. It was suggested in~\cite{Cordova:2017dhq} (see also \cite{Hartman:2016lgu,Delacretaz:2018cfk}) that  $\tau>0$ should hold.
We will show that causality of the CFT in the space-times~\eqref{onelargeintro},\eqref{eq:orig4dppintrod} implies $\tau>0$ for {\it all} operators. It is also enough to assume a bounded from below energy, instead of causality.
Since the space-times~\eqref{onelargeintro},\eqref{eq:orig4dppintrod} have a global timelike Killing vector, there is no ergosphere and no apparent source of instabilities. It is therefore expected that quantum fields on such plane waves behave causally.\footnote{Locality is a key ingredient in the argument: non-local counterexamples can be constructed by considering free fields with spin $\geq 2$, which, however, require higher-derivative kinetic terms. We thank Leonardo Rastelli for raising this point. Furthermore notice that  we will quantize free and weakly-coupled fields on this spacetime and observe that causality is obeyed and the energy is positive. See also~\cite{Gibbons:1975jb} for an early discussion of free fields on such stationary but non-static space-times.} 

\subsection{Applications for Partition Functions} 

Recall that, if we ignore the dependence of spectrum on angular momentum,  the asymptotic growth of the density of operators in 1+1 dimensions is controlled by the Cardy formula~\cite{Cardy:1986ie}, and, more generally in $d + 1$ space-time dimensions, we find (see for instance \cite{Bhattacharyya:2007vs,Shaghoulian:2015lcn,Benjamin:2023qsc,Benjamin:2024kdg})
\begin{equation}\label{eq:Cardyd}
\log \rho(\Delta) \sim f\, \Delta^{d \over d + 1}\,, \qquad \Delta \gg 1
\end{equation}
where \(f\) is the thermal free energy density of the CFT. 

Here we will be interested in obtaining $\log\rho(\tau,\{J_i\})$, in the regime of large angular momentum in the scaling limits~\eqref{scalingone},~\eqref{scalingtwo}.
To that end, we consider 
the following partition function in 2+1 dimensions\footnote{This approach is easily generalized to any other dimensions with only one large angular momentum.} 
\begin{equation}\label{eq:partitionfunctioneps3d}
    Z_{3d}(\beta,\epsilon)
    = \sum_{\mathcal O} e^{-\beta\bigl(\Delta_\mathcal{O}-(1-\epsilon)J_{z\mathcal{O}}\bigr)}
    = \sum_{\mathcal O} e^{-\beta \tau_\mathcal{O} - \beta \epsilon J_{z\mathcal{O}}}\,,
\end{equation}
and in 3+1 dimensions we consider 
\begin{equation}\label{eq:partitionfunctioneps4d}
    Z_{4d}(\beta,\epsilon_{1,2})
    = \sum_{\mathcal O} e^{-\beta\bigl(\Delta_\mathcal{O}-(1-\epsilon_1)J_{\mathcal{O}1}-(1-\epsilon_2)J_{\mathcal{O}2}\bigr)}
    = \sum_{\mathcal O} e^{-\beta \tau_\mathcal{O} - \beta \epsilon_1 J_{\mathcal{O}1}-\beta \epsilon_2 J_{\mathcal{O}2}}\,,
\end{equation}
The above partition functions are well defined for $\epsilon,\epsilon_{1,2}>0$. The regime of large angular momentum in the scaling limits~\eqref{scalingone},\eqref{scalingtwo} arises upon taking $\epsilon,\epsilon_{1,2}\to0^+$ (with fixed $\epsilon_1/\epsilon_2$).

Recently, the limit $\epsilon\to0^+$ and $\epsilon_{1,2}\to0^+$ with fixed $\epsilon_1/\epsilon_2$ was  investigated in~\cite{SemiUniversality} using an expansion around small $\beta$, that led to a proposal that 
\begin{align}
    \log Z_{3d}(\beta,\epsilon)&=\frac{2\pi \beta}{ \epsilon }\,\mathcal F_{3d}(\beta)+\mathcal{O}(\epsilon^0)~, \label{eq:semiuniversality1}\\
     \log Z_{4d}(\beta,\epsilon_{1,2})&= \frac{\beta \pi^2}{2\epsilon_1\epsilon_2 }\, \mathcal F_{4d}(\beta)+\mathcal{O}(\epsilon^0)\label{eq:semiuniversality2}
\end{align}
We see that in 2+1 dimensions, $2\pi/\epsilon$ behaves as an effective ``emergent'' volume and the partition function has a form reminiscent of an extensive thermodynamic system. Similarly, in 3+1 dimensions, $\frac{\pi^2}{2\epsilon_1\epsilon_2}$ plays the role of an effective volume.\footnote{In supersymmetric theories, {\it complex} angular momentum is sometimes used as an infrared cutoff on the volume of space~\cite{Lossev:1997bz,Nekrasov:2002qd}. Here, $\epsilon\to0^+$ and $\epsilon_{1,2}\to0^+$ correspond to large, real, angular momentum limits which lead to an effective divergent volume.} Let us emphasize that the equations above were shown to hold at any order in the high temperature expansion in \cite{SemiUniversality}.

Another point of view (see also~\cite{SemiUniversality}) on the extensive potentials at large angular momentum is to consider the density of states, which can be found easily from~\eqref{eq:semiuniversality1},\eqref{eq:semiuniversality2}. 
We obtain
\begin{align} \label{eq:proposalsemiuni1}
    &\log \rho_{3d}(\tau,J_z) \sim \sqrt{J_z} \, G_{3d}\left(\frac{\tau}{\sqrt{J_z}}\right) \,, 
    && J_z \to \infty \,, \quad \frac{\tau}{\sqrt{J_z}}= \text{fixed} \,; \\ 
\label{eq:proposalsemiuni2}
    &\log \rho_{4d}(\tau,J_{1,2}) \sim (J_1J_2)^{1/3} \, G_{4d}\left(\frac{\tau}{(J_1J_2)^{1/3}}\right) \,, 
    && J_{1,2}\to \infty \,, \quad \frac{J_1}{J_2}~,\frac{\tau}{(J_1J_2)^{1/3}}= \text{fixed} \ ,
\end{align}
where $G_{3d/4d}$ and $\mathcal F_{3d/4d}$ are related by Laplace transforms.
One should compare this to the thermodynamics of extensive systems, where the entropy obeys $S(E,V)=Vs(E/V)$. We see that $\sqrt J_z$ and $(J_1J_2)^{1/3}$ are proportional to the ``emergent volume.''

We can use the weakly-coupled partons~\eqref{FOCKstatesintro},\eqref{FOCKstatesintro2} to calculate the thermodynamic potentials at low temperature $\beta\gg1$, or equivalently, at $\tau\ll \sqrt{J_z}$, and  $\tau \ll (J_1J_2)^{1/3}$. 
We obtain the following results: \begin{align}
&\mathcal F_{3d}(\beta) \sim {1\over 2\pi \beta^2 }e^{-\beta \tau_{\min}} \,,
    \quad 
    &&G_{3d}\!\left(\frac{\tau}{\sqrt{J_z}}\right) \sim -\,\frac{2}{\tau_{\min}}\,
    \frac{\tau}{\sqrt{J_z}}\,
    \log\!\left(\frac{\tau}{\tau_{\min}\sqrt{J_z}}\right)\,\notag,\\
&\mathcal F_{4d}(\beta) \sim {1\over 2\pi^2 \beta^3 }e^{-\beta \tau_{\min}} \,,
    \quad 
    &&G_{4d}\!\left(\frac{\tau}{(J_1J_2)^{1/3}}\right) \sim -\frac{3}{\tau_{\min}}{\tau\over (J_1J_2)^{1/3}}
\log\!\left(\frac{\tau}{\tau_{\min}\,(J_1J_2)^{1/3}}\right)~, \notag
\end{align}
where $\tau_{\min}$ is the twist gap of the CFT.
This type of entropy density is characteristic  of a free gas of gapped particles, which indeed is the situation in the lightcone bootstrap regime. 

This extensive thermodynamics follows directly from the fact that the corresponding large spin operators are captured by the pp-wave geometries~\eqref{onelargeintro},\eqref{eq:orig4dppintrod}. The pp wave geometries have infinite spatial volumes and we expect partition functions to be proportional to the volume. The precise dictionary is as follows. For the geometries corresponding to a single large angular momentum we have in $d+1$ space-time dimensions
\begin{equation} 
\text{Vol}_d\longleftrightarrow \frac{\pi^\frac{d+1}{2}}{\epsilon \Gamma(d/2+1/2)}~, \end{equation}
while for the 3+1 dimensional geometry corresponding to 
$J_{1,2}\to \infty$ with fixed $\frac{J_1}{J_2}$  we obtain 
\begin{equation} 
\text{Vol}_3\longleftrightarrow {\pi^2\over 2\epsilon_1\epsilon_2}.
\end{equation} 
Regularizing the spatial volume of the pp-wave geometries in this fashion,  we obtain exact agreement with \eqref{eq:semiuniversality1}, and \eqref{eq:semiuniversality2}.\footnote{We have to clarify what we mean by spatial volume. 
In stationary but non-static space-times, there is no preferred space-like slice. 
However, we can introduce a canonical volume form, by examining the definition of energy in this geometry as explained in Section \ref{sec:staticdigression}.%
}
The locality of the quantum fields on the pp-wave geometries gives a proof of the universal scalings in the large spin limits.

In the high temperature limit locality implies that  $\mathcal  F(\beta) \sim \frac{f}{\beta^{d-1}}$. While the low temperature limit is dominated by low twist operators, the high temperature limit is controlled by the thermal effective action \cite{Banerjee:2012iz,Benjamin:2023qsc,Benjamin:2024kdg} (which can be recast in the language of the thermal bootstrap \cite{Iliesiu:2018fao,Iliesiu:2018zlz,Marchetto:2023xap,Barrat:2024fwq,Cho:2024kxn,Barrat:2025nvu}).

\subsection{Open Questions and Outline}
\begin{itemize}
\item[$\star$] Can we use the pp-wave geometry to learn about nontrivial correlation functions in the large spin limit?
For instance, correlation functions of the form $\langle J,\tau|\,\mathcal O_1\,\mathcal O_2\,\cdots|J,\tau\rangle$. See e.g.\ \cite{El-Showk:2011yvt,Parnachev:2020fna,Marchetto:2023fcw} for such calculations in a different context. 
What are the implications of the Heisenberg symmetry in the large spin limit for such correlation functions?  
\item [$\star$] Are there examples where $\mathcal F_{3d,4d}(\beta)$ is non-analytic and exhibits phase transitions? Note that this is entirely possible since we are effectively at large volume in the large spin limit (see also figure~\ref{analyticity}).\footnote{In this work we only perform calculations in free-theories and weakly coupled example for which $\mathcal F_{3d/4d}(\beta)$ is analytic at any temperature. However $\mathcal N = 4$ SYM at large $N$ is believed to have three different phases: see \cite{Kim:2023sig,Choi:2024xnv,Bajaj:2024utv,SemiUniversality}. We are not aware of any finite $N$ theory with non-analytic $\mathcal F_{3d,4d}$.} 

\item[$\star$] Is it possible to rigorously establish properties such as causality and bounded from below energy for CFTs placed in the pp wave geometries? 

\item[$\star$] We will see that the spectrum of free fermions and free bosons on the pp-wave~\eqref{onelargeintro} coincide. This follows from the fact that certain supercharges commute with $\Delta-J_z$. Therefore, the large spin limit $J_z\to\infty$ at fixed $\tau/\sqrt{J_z}$ preserves supersymmetry: it would be interesting to analyze it further. In particular, this perspective may  give new understandings of various large charge/large spin limits of the supersymmetric indices, see~\cite{DiPietro:2014bca,Beem:2017ooy,ArabiArdehali:2019tdm,Choi:2018hmj,Cabo-Bizet:2019osg,Cassani:2021fyv,
Rastelli:2023sfk} for a very partial list of references on this topic.

\item[$\star$] For large $N$ theories, it would be interesting
to study gravitational backgrounds with a pp-wave boundary~\eqref{onelargeintro},\eqref{eq:orig4dppintrod}. Kerr black holes with these properties were considered in~\cite{Maldacena:2008wh}.

\end{itemize}

The paper is organized as follows. In Section~\ref{formlowtohightwist} we study the spectrum of spinning operators in CFTs by analyzing the partition functions~\eqref{eq:partitionfunctioneps3d},\eqref{eq:partitionfunctioneps4d} in the large-spin regime, both at low and at high temperature. In Section~\ref{eq:universalsing} we explain how to obtain the pp wave  geometries in $2+1$ and $3+1$ dimensions and analyze their symmetries. Finally, in Section~\ref{examples} we present a number of explicit examples, including free scalars and fermions, Maxwell theory, and the Ising model in the $\varepsilon$-expansion to first order. In Appendix~\ref{app:ckv}, we present the complete list of conformal Killing vectors for the metric \eqref{onelargeintro}.

\section{From Small to Large Twist}\label{formlowtohightwist}

CFTs %
are generally strongly coupled and relatively little is known about them. However, in various limit such as large charge or large spin, there are interesting results about the spectrum of operators. We concentrate on large spin operators, which is a universal sub-sector.\footnote{Very interesting results also exist in two spacetime dimensions, see~\cite{Pal:2022vqc,Pal:2025yvz}, but we do not discuss this case here.} 
For simplicity consider 2+1 dimensional theories and define the usual twist \begin{equation} 
\label{twistdefsec}\tau=\Delta-J_z~, \quad J_z \in \frac12 \mathbb{Z}\end{equation}
with $\Delta$ the scaling dimension and $J_z$ the spin of the operator with respect to rotations around the $z$ axis. The unitarity bounds on  local primary operator (excluding the identity) are~\cite{Poland:2018epd} 
\begin{align*}
\Delta \geq 
\left\{ \begin{array}{ll} 
\frac12 & J = 0 \\ 
1 & J = 1/2 \\ 
J + 1 & J \ge 1 
\end{array} \right.
\end{align*}
In particular, since $|J_z|\leq J$, we see that the twist~\eqref{twistdefsec} satisfies $\tau\geq \frac12$. 

We can ask what are the operators with the lowest possible values of $\tau$ in generic interacting CFTs. 
It turns out that operators with finite $\tau$ but large $J_z$ are weakly interacting and furnish a Fock space, see~\cite{Komargodski:2012ek,Fitzpatrick:2012yx} as well as~\cite{Fardelli:2025eun} for more recent developments and references.

For brevity, let $\partial_\pm = \partial_x \pm i \partial_y$ denote a derivative with $\Delta =1$ and $J_z=\pm1$. 
Consider any primary operator $\cal O$ with $J_z=J_0$ and scaling dimension $\Delta=\Delta_0$ and consider general multi-twist operators of the form 
\begin{equation}\label{FOCKstates}\partial_+^{m_1}{\cal O}\cdots \partial_+^{m_n}{\cal O}~.\end{equation}
Classically these operators have $\Delta=n\Delta_0+\sum_{i=1}^n m_i$ and angular momentum $J_z=nJ_0+\sum_{i=1}^n m_i$ and hence twist $\tau = n\tau_0$. 

Usually composite operators in CFT receive large quantum corrections, and naive estimation fails. For example, the dimension of a composite operator is not equal to the sum of the dimensions of its constituents.
However, for $m_i\gg n$ it turns out that the  composite operators behave almost classically and the twist just adds up due to the constituents ,``partons'', i.e. $\tau = n \tau_0$ is correct to a good approximation.
The deviations from the classical predictions scale as $\sim m_i^{-1}$; we will be more precise about this below. 

Due to this Fock space of weakly-coupled composite operators, the partition function \begin{equation} Z=\sum e^{-\beta(\Delta-J_z)}\end{equation} diverges. Indeed, the divergence is due to the fact that choosing arbitrary large $\{m_i\}$ does not affect the twist spectrum to the leading order.\footnote{
This is of course a similar divergence to the one we encounter in flat space QFT if we try to calculate \begin{equation} \sum e^{-\beta(H-P)}~,\end{equation} 
which diverges from highly-boosted multi-particle states, though the details are quite different.}   
This divergence can be regularized by adding $e^{-\beta \epsilon J_z}$ to the partition function, with $\epsilon>0$. Thus we consider the regularized partition function
\begin{equation}\label{regularizedZ} Z=\sum e^{-\beta(\Delta-J_z)-\beta\epsilon J_z}~.
\end{equation} 
At $\beta\gg1$ the partition function contains information about the Fock states~\eqref{FOCKstates}. As we increase the temperature, states with larger and larger twist become important and the Fock space picture breaks down at sufficiently large twist. For Fock states as in~\eqref{FOCKstates}
the corrections to the leading order formula take the following form $\Delta=n\Delta_0+\sum_{i=1}^n m_i+\delta \Delta$ with $\delta\Delta \sim{n^2\over m_i}\sim {n^3\over J_z}$ where we assumed that typically $m_i\sim J_z/n$. Therefore $\tau-n\tau_0\sim {n^3\over J_z}$. The corrections overwhelm the classical prediction for $n^2/J_z\sim 1$, which means that when 
\begin{equation}\label{strongcoupling}\tau\sim \sqrt J_z\end{equation} 
the weakly-coupled partons picture breaks down.\footnote{\label{fn:precise}{This counting is essentially identical to the argument regarding when the charged Fock space breaks down in~\cite{Cuomo:2022kio}. The Fock space effective theory can break down earlier, in some cases. To estimate that the corrections behave as $\delta \Delta\sim n^3/J_z$, we have estimated the sum over pairs  contributes $n^2$ and the typical inter-parton distance being $J_z/n$. This is  too naive as it ignores the rotation of the partons. The correct calculation, which leads to the same conclusion, is presented in~\cite{Fardelli:2024heb,Kravchuk:2024wmv}. It turns out that when only one of the spin is taken to be large the breakdown of the EFT is always expected for $\tau \sim \sqrt{J}$, independently on the spacetime dimensions. }} This will imply that at temperatures $\beta\lesssim1$ in~\eqref{regularizedZ}, we will not be able to trust the weakly coupled partons picture. 

Recently, in~\cite{SemiUniversality}  the authors proposed a conjecture that the universal behavior of the partition function~\eqref{regularizedZ} should behave as 
\begin{equation}\label{eq:semiuniversality}
 \log Z(\beta,\epsilon)\sim \frac{2\pi \beta}{ \epsilon}\, \mathcal F_{3d}(\beta)+\mathcal{O}(\epsilon^0),\quad \epsilon\to 0^+,
\end{equation}
this statement at very high temperature $\beta\to0$ can be seen from the high temperature effective action~\cite{Benjamin:2023qsc} as shown to all orders in~\cite{SemiUniversality}. In that limit we obtain $\mathcal F(\beta)={f\over \beta^3}$, where $f>0$ is some model-dependent coefficient (coinciding with the coefficient that appears in the flat space energy-density). 

\begin{figure}[t]
\centering
\begin{tikzpicture}[scale=1.1,>=latex]

  \def\xmin{0}
  \def\xmax{5}   %
  \def\ymin{0}
  \def\ymax{4.2} %

\fill[gray!10] (\xmin,\ymin) rectangle (\xmax,\ymax);

\draw[thick] (\xmin,\ymin) -- (\xmax,\ymin); %
\draw[thick] (\xmin,\ymin) -- (\xmin,\ymax); %
\draw[thick] (\xmax,\ymin) -- (\xmax,\ymax); %

  \draw[thick] (\xmin,\ymin) -- (\xmax,\ymin);
\node[below] at (\xmax+0.30,\ymin+0.15) {$\epsilon$};

   \draw[->,thick] (\xmin,\ymin) -- (\xmin,\ymax+0.6) node[left] {$1/\beta$};

  \foreach \e/\x in {0/0,1/2.5,2/5}{
    \draw[thick] (\x,\ymin) -- (\x,\ymin-0.12);
    \node[below] at (\x,\ymin-0.12) {$\e$};
  }

  \node[align=center] at (0.5*\xmax,0.55*\ymax)
  {$\log Z(\beta,\epsilon)$ is real analytic\\
   };

\end{tikzpicture}
\caption{Half-infinite slab: At $\epsilon=0$ and $\epsilon =2$,  $\log Z$ develops a pole and the residue of the pole, as a function of $\beta$ may be non-analytic.}\label{analyticity}
\end{figure}
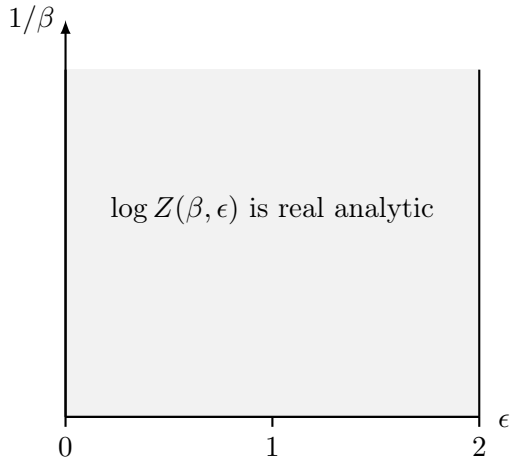

We can verify~\eqref{eq:semiuniversality} at low temperatures from the high-spin Fock space as follows (see also~\cite{SemiUniversality}). 
Let us start from some primary $\cal O$ of twist $\tau_0$ and spin $J_z=J_0$ and consider the descendants  $\partial_+^m{\cal O}$. For these states we have the one-particle partition function $$Z_1=e^{-\beta \tau_0-\beta\epsilon J_0}\sum_{m}e^{-\beta\epsilon m }={e^{-\beta \tau_0} \over \beta \epsilon} +\cdots~,$$
where we expanded to leading order in the small-$\epsilon$ expansion. 
We then build a Fock space out of these fields via the plethystic exponential. Keeping only the leading term at low temperatures we obtain 
\begin{equation}\label{pla}Z=e^{{1\over \beta \epsilon} e^{-\beta \tau_0}+\cdots} ~.\end{equation}
We see that this is compatible with~\eqref{eq:semiuniversality}. 
Since the above calculation at low temperatures is valid for any primary operator, the true low temperature limit is dominated by the minimal twist operator (other than the identity) with twist $\tau_{\rm min}$.
\begin{figure}
\centering
    \includegraphics[scale = 0.28]{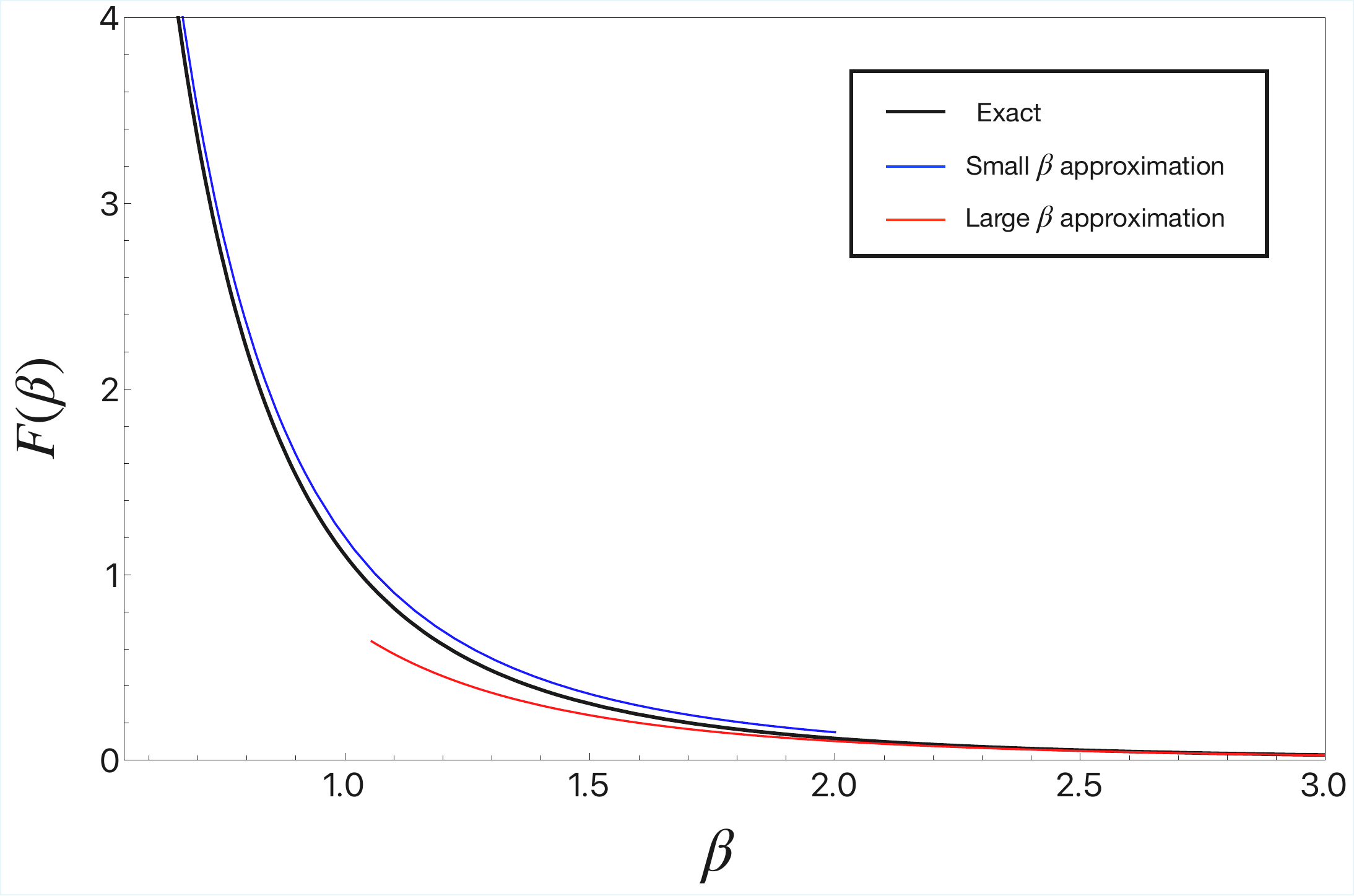}
    \caption{Exact $F_{3d}(\beta)$ for the free scalar theory (black) compared with its low-temperature (red) and high-temperature (blue) asymptotics. 
    We emphasize that this plot is meant only as an illustrative free-theory example: in interacting CFTs the interpolation between the two regimes can be more intricate, and may involve additional structure---or even phase transitions.
}
    \label{fig:FFreeScalar}
\end{figure}
We summarize that
\[\begin{cases}
 \mathcal F_{3d}(\beta)= \frac{e^{-\beta\tau_{\rm min}}}{2 \pi \beta^2}, & \beta\to \infty\\
 \mathcal F_{3d}(\beta)={f\over {2 \pi\beta^3}},
 & \beta\to 0~.
\end{cases}
\]
As an illustrative example, the case of the free scalar theory is presented in Fig.\ref{fig:FFreeScalar}.
There could be phase transitions between the low temperature and the high temperature regimes. Indeed, while the partition function is analytic for  $0<\epsilon<2$, the residue of $
\log Z$ at the boundary of this domain could fail to be analytic: see Fig.~\ref{analyticity}.

We can directly calculate the density of states $\rho(\tau,J_z)$ via the  Legendre transform:
\begin{equation}\log \rho(\tau,J_z) = {2 \pi \beta \over \epsilon}\mathcal F(\beta)+ \beta\tau+\beta\epsilon J_z ~.\end{equation} 
From that we find that the small $\epsilon$, fixed $\beta$ limit corresponds to%
\begin{equation} 
J_z\to\infty~,\quad {\tau\over \sqrt J_z} ={\rm fixed}~.
\end{equation}
From~\eqref{eq:semiuniversality} it follows that the micro-canonical density in this limit takes the form 
\begin{equation}\label{micro}
\log \rho(\tau,J_z)= \sqrt J_z  G_{3d}\left({\tau\over\sqrt J_z}\right)~.  \end{equation}
Our general low temperatures prediction 
$\log Z \sim {e^{-\beta\tau_{\rm min}}\over \beta\epsilon}$ then corresponds to 
\begin{equation}\label{higherspindensity}\log \rho (\tau, J_z) =\sqrt{J_z}   
{2\tau\over \tau_{\rm min}\sqrt{J_z}}\log\left({\tau_{\rm min} \sqrt J_z\over \tau}\right)~,\end{equation}
which is valid for  $J_z\to\infty$ and $\tau/\sqrt{J_z}$ fixed and much smaller than 1. 
The formula~\eqref{higherspindensity} counts the low-twist high spin operators in the Fock space.

Performing the Legendre transform at high temperature we find 
\begin{equation}\label{higherspindensityi}\log \rho (\tau, J_z) =3\sqrt{J_z}   
f^{1/3}\left(\tau\over {\sqrt J_z}\right)^{1/3}~.\end{equation}
The structure~\eqref{micro} suggests a macroscopic limit in the large spin limit, identifying the volume with $\sim \sqrt J_z$. The low temperature entropy density we obtained above is characteristic to any local theory with a unique gapped ground state and dilute, weakly interacting excitations. Both for small and large energy densities, the entropy density that we found is monotonic and concave, as required.

In summary, we see that there is emergent extensive physics in the large spin limit~\eqref{micro}. We have shown that at small energy densities we can obtain concrete predictions from the large-spin effective theory~\eqref{higherspindensity}. At  energy density of order 1 we hit strong coupling for the large  -spin Fock space~\eqref{strongcoupling} and the physics changes. 

Phase transitions may occur as we dial $\beta$. It is not necessary to take a large $N$ limit for that, since the systems effectively develops a large volume limit at large $J_z$. 

\subsection{3+1 Dimensions} 

The kinematics in 3+1 dimensions are different so we show explicitly the main differences here. We consider the twist \begin{equation}\label{twist4d}\tau=\Delta-J_{1}-J_{2}~,\end{equation}
where $J_{1,2}$ correspond to the rotations in the planes $12$ and $34$ in  $\mathbb{R}^4$, respectively. It is common to work in an alternative basis in which the \(SO(4)\) symmetry is written as \(SU(2)_L \times SU(2)_R\). In this basis, one introduces spins \(s_1, s_2 \in \frac12 \mathbb{N}_0\), corresponding to the representations under \(SU(2)_L\) and \(SU(2)_R\), respectively.
Denoting the Cartan values of $SU_{L,R}(2)$ by $s_{1z},s_{2z}$, we have $J_1=s_{1z}+s_{2z}$ and 
$J_2=s_{1z}-s_{2z}$. The unitarity bounds for primary operators 
\[
\begin{array}{ll}
s_1 =s_2=0: & \Delta \ge 1 \quad \text{(except the identity with }\Delta=0),\\[4pt]
s_1=0,\ s_2>0\ \text{or}\ s_2=0,\ s_1>0: & \Delta \ge s_1+s_2+1,\\[4pt]
s_1>0,\ s_2>0: & \Delta \ge s_1+s_2+2.
\end{array}~,
\]
note that it does not follow from the known unitarity bounds that $\tau$ in~\eqref{twist4d} is non-negative. However, it is in fact non-negative in all known examples~\cite{Cordova:2017dhq} (see also \cite{Hartman:2016lgu,Delacretaz:2018cfk}).

Denoting derivatives with $\Delta=J_1=1$ by $\partial^{(1)}_{+} $ and derivatives with $\Delta=J_2=1$ with $\partial^{(2)}_{+} $, we have the Fock-space of higher-spin operators 
\begin{equation}\label{FOCKstates4d}
\partial_{+}^{(1) p_1}\partial_{+}^{(2) q_1}{\cal O}\cdots \partial_{+}^{(1)p_n}\partial_{+}^{(2) q_n}{\cal O}~.
\end{equation}
Classically these operators have $\Delta=n\Delta_0+\sum_{i=1}^n (p_i+q_i)$ and  twist $\tau = n\tau_0$. 
Assuming the $p_i$ and $q_i$ are of the same order $\sim J/n$, the interactions between the partons can be estimated to scale as $\tau-n\tau_0\sim n^4/J_1J_2$ and hence these corrections overwhelm the classical approximation for 
\begin{equation}\tau\sim (J_1J_2)^{1/3}~.\end{equation}
This is where the large-spin effective theory breaks down.\footnote{As in footnote~\ref{fn:precise}, the estimate can be obtained by a naive sum over parton pairs, but the correct derivation is in~\cite{Fardelli:2024heb,Kravchuk:2024wmv}.} 

We consider the regularized partition function 
\begin{equation}Z=\sum e^{-\beta\tau-\beta \epsilon_1J_1-\beta \epsilon_2J_2} ~.\end{equation}
We can obtain the low-temperature prediction for this partition function by summing over the states in the Fock space~\eqref{FOCKstates4d}. For the one particle states we find $Z_1= {1\over \beta^2\epsilon_1\epsilon_2}e^{-\beta\tau_0}$ and hence using the plethystic exponential we obtain the low temperature limit
\begin{equation}Z=e^{{1\over \beta^2\epsilon_1\epsilon_2} e^{-\beta\tau_{\rm min}}}~.\end{equation}
At high temperature we obtain $\log Z = {f\over \beta^3\epsilon_1\epsilon_2}$. The analog of~\eqref{eq:semiuniversality} is thus
\begin{equation}\label{4duniversality}
\log Z = {\pi^2 \beta \over 2\epsilon_1\epsilon_2} \mathcal F_{4d}(\beta)~,\quad \log\rho(\tau,J_1,J_2)=(J_1J_2)^{1/3}G_{4d}\left({\tau\over (J_1J_2)^{1/3} }\right)~,
\end{equation}
in the canonical and micro-canonical ensembles, respectively.  The entropy at low energy density is again like a system of gapped weakly interacting excitations, while at high temperature we now find entropy characteristic of a local 3+1 dimensional system.

\section{Conformal Fields on Plane Waves}\label{eq:universalsing}
Here we derive the geometries that reproduce the equation of state in the large-spin limit. That is, we provide a geometry that explains the emergent extensive thermodynamics~\eqref{micro} and its counterpart in 3+1 dimensions~\eqref{4duniversality}.

\subsection{Thermodynamics in stationary but non-static spacetimes}
\label{sec:staticdigression}

Before specializing to our explicit backgrounds, let us emphasize a general point. The effective metrics we will encounter
are stationary but non-static. In particular, although there exists a global timelike Killing vector
\(\xi=\partial_t\) generating time translations, we cannot synchronize clocks in a non-static spacetime. As a result, constant-\(t\) slices are not canonically preferred and there is no unique, diffeomorphism-invariant notion of a ``spatial slice'' and hence we have to explain what ``spatial volume'' means.
Let us be more concrete: since the background admits a timelike Killing vector $\xi=\partial_t$, the most general stationary metric can be written in Kaluza--Klein form as
\begin{equation}\label{eq:generalmetric}
ds^2 = -N^2(x)\left(dt+A_i(x)dx^i\right)^2 + h_{ij}(x)\,dx^i dx^j ,
\end{equation}
where $N$ is the referred to as ``lapse function'',
$A_i(x)$ is a spatial one-form encoding the non-staticity of the geometry, and $h_{ij}$ is a positive-definite metric. If $A_i(x)$ is not a pure gauge, then the constant-$t$ hypersurfaces are not orthogonal to the Killing flow.

Nevertheless, thermodynamics is defined unambiguously in the Hamiltonian approach. Given a conserved and symmetric stress
tensor \(T_{\mu\nu}\) and a Killing vector \(\xi^\mu\) define a conserved charge
\begin{equation}
  E \;\equiv\; \int_{\Sigma} d\Sigma_\mu\, T^{\mu}{}_{\nu}\,\xi^\nu\,,
  \label{eq:H_xi}
\end{equation}
where \(\Sigma\) is {\it any space} like surface without boundaries. 

 Equation \eqref{eq:H_xi}, when $\xi^\nu = \delta^{\nu \, 0}$, reduces to 
    \begin{equation}
        E = \int T^0_{\, 0}\, N \sqrt{\det h}\, d^{d-1}x ,
    \end{equation}
    where $h_{ij}$ is precisely the Kaluza--Klein metric.
 We emphasize that $h_{ij}(x)$ is not associated with any special slice of the metric: it is instead the metric on the space of orbits of the timelike Killing vector. However, it is also the natural metric that defines the thermodynamics of the space \eqref{eq:generalmetric}(see also \cite{landau2004theoretical}).%
 The volume form \begin{equation} N \sqrt{\det h}\, d^{d-1}x\end{equation} is explicitly gauge invariant under  diffeomorphisms of $x^i$ and also KK gauge transformations $t\to t+\Lambda(x)$, $A_i\to A_i-\partial_i\Lambda(x)$.

We have now identified a notion of volume, which remains meaningful even in non-static spacetimes\footnote{It would be interesting to derive it also from the perspective of~\cite{Bah:2025zzk}}. 
The next issue is to understand extensivity in such a setting. 

At high temperature the situation is clear: one way to see this is to note that, in the thermal effective action, the leading contribution is extensive. 
At low temperature, however, it is in general much harder to make definitive statements. 
In our case we will be in a favorable situation. 
Indeed, the presence of an enlarged isometry group will allow us to unambiguously isolate the ``volume'' contributions at any temperature.

\subsection{2+1 d}
Let us consider the Lorentzian cylinder 
\begin{gather}
    ds^2 = -dt^2 + d\theta^2 + \sin^2 \theta d\phi^2,
\end{gather}
and transform to a rotating reference frame, rotating slightly below the speed of light at the equator ($\epsilon>0$):
 $\phi \to \phi + (1-\epsilon) t$. The metric becomes 
\begin{gather}
    ds^2 = -dt^2 + d\theta^2 + \sin^2 \theta (d\phi + (1-\epsilon) dt)^2,
\end{gather}
Now we need to perform a scaling limit which will allow us to extract the leading singularity as $\epsilon\to0$. 

To explain the correct scaling limit we digress for a moment and discuss the wave functions of particles with large $J_z = J$ on $S^2$. 
A convenient starting point is the highest-weight spherical harmonic,
whose angular dependence can be written (up to an overall normalization) as
\begin{equation}
    \Psi_0(\theta,\phi)\ \propto\ e^{iJ\phi}\,\sin^{J}\theta\ .
\end{equation}
This is sharply localized near the equator with  $\delta\theta \sim J^{-\frac{1}{2}}$. A more common approach in the literature is to transform the Lorentzian cylinder to AdS$_3$ of radius $L$, where spinning particles are at distance $D\sim {L\over 2} \log J_z $ from the center of AdS$_3$. These are of course equivalent approaches since the two spaces are related by a Weyl rescaling. Another perspective, discussed in Ref.~\cite{SemiUniversality}, is that the integrals appearing in thermodynamics become saturated in the vicinity of the equator, reflecting the localization of the wave function in this region.

Thus, using the above intuition, we rescale the coordinates as  follows
\begin{gather}\label{scalimit}
    \phi = \epsilon  x, \quad \theta =  \frac{\pi}{2} + \sqrt{ 2 \epsilon } y~.
\end{gather}
This is the same as Penrose's expansion around null geodesics \cite{Penrose1976,BlauEtAlPenrose2002}.
Then expanding in $\epsilon$ the metric becomes
\begin{gather}
    ds^2 
    = 2 \epsilon\left(-(1 + y^2) dt^2 + dy^2 +  dt dx\right) + \mathcal{O}(\epsilon^2)~.
\end{gather}
We can Weyl rescale away the leading factor $\epsilon$, that allow to consider the limit $\epsilon\to 0$,  and get the following metric 
\begin{gather}\label{ppBrink}
      ds^2 =-(1 + y^2) dt^2 +  dt dx + dy^2~.
\end{gather}
This is a Brinkmann pp-wave with $\partial_x$ a covariantly constant null Killing vector. 
 The metric in equation \eqref{ppBrink}
admits a three-dimensional Heisenberg isometry group \(\mathbb H_3\), generated by the Killing vectors
\[
K_1=\partial_x,\qquad
K_2=\cos t\,\partial_y + y\sin t\,\partial_x,\qquad
K_3=\sin t\,\partial_y - y\cos t\,\partial_x,
\]
 Defining $K_\pm = K_1\pm i K_3$ we observe that $K_+$ annihilate the vaccum and in fact $K_\pm$ acts as creation/annihilation operators. 
The generators above  satisfy the Heisenberg commutation relations
\[
[K_+,K_-] = 2 i\,K_1,\qquad
[K_1,K_+]=[K_1,K_-]=0.
\]
In addition, the metric is invariant under time translations generated by $T=\partial_t$.\footnote{The emergence of the Heisenberg group in a different null limit was also noted in~\cite{Erramilli:2022kgp}.}
 Note that from the commutations above and the fact that $K_+$ annihilate the vaccum unitarity implies that $K_1$ is positive definite.  We analyze the conformal Killing vectors (CKVs) of~\eqref{ppBrink} in appendix~\ref{app:ckv}. In particular, it follows from the algebra of CKVs that the energy $T$ is non-negative.

 A variant of~\eqref{ppBrink} is obtained by transforming $x=x'+t$, which leads to $g_{00}=-y^2$. In this space, the energy spectrum is nonnegative and always infinitely degenerate. In our coordinates the energy degeneracy is removed and the thermal ensemble is defined with respect to our $\partial_t$. Positivity of the energy together with causality in the spacetime with $g_{00}=-y^2$ implies the positive energy condition in~\eqref{ppBrink}. We can therefore derive the positive energy spectrum in~\eqref{ppBrink} either from the algebra of CKVs or from causality. This is an interesting point of view on the unitarity bounds in CFTs.

Finally, it is interesting to comment that time evolution in~\eqref{ppBrink} is {\it not} equivalent to the spectrum of $\Delta-J_z$. Indeed the former necessarily has infinite degeneracies. Instead, we obtain, as explained in appendix~\ref{app:ckv}, that time evolution in~\eqref{ppBrink} is given by $P_u-K_v+P_v$. The combination $P_u-K_v$ is indeed related to the twist $\Delta-J_z$ while $P_v$ lifts the infinite degeneracy. That is why the pp-wave geometry leads to sensible partition functions.

At finite temperature (i.e.\ after compactifying \(t\sim t+\beta\)), only the subgroup commuting with \(T\) survives.
The residual continuous isometries are the translations along \(t\) and \(x\), generated by \(T\) and \(K_1\), respectively.

To better understand thermodynamics in this space it is convenient to define $y=\sqrt 2 \sinh\rho$, and Weyl rescale the metric by ${1\over 2\cosh^2\rho}$ to obtain 
\begin{gather}\label{ppKK}
      ds^2 = - dt^2 +  {dt dx \over \cosh^2\rho}+  d\rho^2 = - \left(dt-  A\right)^2+ ds^2_{\rm space}~\notag\\
      A = \frac{dx}{2 \cosh^2 \rho}, \quad
      ds_{{\rm space}}^2 =  {1\over 4\cosh^4\rho}dx^2  + d\rho^2,
\end{gather}
where we wrote the metric in the Kaluza-Klein form. This allows us to identify the base space with $ds^2_{\rm space}$, which has a diverging volume due to the non-compact $x$ coordinate. %
Therefore, thermodynamics in this space leads to a partition function of the form 
\begin{equation} 
\log Z= L_x \beta \, \mathcal F(\beta)~,
\end{equation} 
which simply follows from locality. 
All that remains is to identify the diverging volume with the divergence in the high spin limit
$L_x\leftrightarrow {2\pi\over \epsilon}$. The function $\mathcal F(\beta)$ that appears in the thermodynamics on the pp wave is the same as the function $\mathcal F_{3d}(\beta)$ that appeared in the large spin limit.

The universal behavior of the free energy in the large spin limit as well as the extensive behavior of the entropy, are just a consequence of locality in the geometry~\eqref{ppBrink}.

\subsection{3+1d} 
The considerations in 3+1 dimensions are very similar but have an important new ingredient. 
We start from a fast rotating reference frame $\phi_{1,2}\to \phi_{1,2}+(1 - \epsilon_{1,2})t$. 
We get that the metric has the following form
\begin{gather}
    ds^2 =  -dt^2 + d\theta^2 + \sin^2\theta (d\phi_1 + (1-\epsilon_1) dt)^2 + \cos^2 \theta (d\phi_2 + (1- \epsilon_2) dt)^2~.
\end{gather}

In 2+1 dimensions the fast spinning particles were all concentrated near $\theta=\pi/2$ which led to the Penrose scaling limit~\eqref{scalimit}. Here, by contrast, the trajectory is nearly null for all $\theta$. This corresponds to the fact that the Fock space operators~\eqref{FOCKstates4d} wave functions are localized at different $\theta$ depending on $p_i/q_i$.
We are therefore not allowed to zoom on any particular value of $\theta$ since that would miss out on most of the high-spin operators. { Therefore we need to consider a generalized Penrose-type limit which focuses on a family of geodesics, instead of a single geodesic. }

We consider the following rescaling:
\begin{gather}
    d\theta = \sqrt{\gamma(\theta)} dy,  \quad \gamma(\theta) = (\epsilon_1+\epsilon_2)+(\epsilon_1-\epsilon_2) \cos \left(2\theta\right), \quad  \gamma(y) = \gamma(\theta(y))\\
    \quad 
     \phi_1 = \sqrt{ \gamma(y)}\tan\left(\theta(y)\right) u +  \gamma(y) v,\quad \phi_2 = -\sqrt{ \gamma(y)}\cot\left(\theta(y)\right) u +  \gamma(y) v
\end{gather}
 In these rescaled coordinates to leading order in $\epsilon$ we obtain 
\begin{gather}
    ds^2 = \gamma(\theta)(- dt^2 -  2 dt dv - 4 u dt dy +  du^2 + dy^2) + \mathcal{O}(\epsilon_{1,2}^2)
\end{gather}
Dropping the higher order terms and performing a Weyl rescaling we obtain our final metric 
\begin{gather} \label{eq:orig4dpp}
ds^2 = -dt^2 -  2 dt dv - 4 u dt dy + du^2 + dy^2~.
\end{gather}
As before, to understand the thermodynamics it is useful to switch to a KK form and define\footnote{Note, that this transformation is allowed since it does not change the notion of energy, on the other hand the transformation of the form $v' = t + v$ does change this notion.} $t'=t+v$  to obtain 
\begin{gather}\label{4dpp}
ds^2 = - (dt' + 2 u\, dy )^2 + (dv +  2 u\, dy)^2 +  du^2 + dy^2~.
\end{gather}
The base is the standard left-invariant metric on the Heisenberg group $\mathbb{H}_3$, which emerges in the large spin limit. Explicitly, the isometries are $v\to v +a_v, y \to y +a_y$ and the following isometry 
\begin{equation}
u \to u+a_u,\quad  v \to v-2a_u y, \quad  t \to t - 2 a_u y.
\end{equation}
Finally, we have time translations, $t\to t+a_t$.
They obey the usual Heisenberg relation $[U,Y]=2V-2T$, where $Y,U,V,T$ are the Killing vectors implementing the symmetries generated by $a_{y,u,v,t}$, respectively.
{ We also have an extra $SO(2)$ symmetry which rotates the generators $U$ and $V$. Finally we also have time translations.The full isometry group of this metric is therefore given by $\mathbb{R} \times (\mathbb H_3 \rtimes SO(2))$.}
All the other commutators vanish. 
We will comment on the CKVs later in this section. 

Note that since all the generators of the Heisenberg group $\mathbb H_3$ commutes with $T$, it means that $\mathbb H_3$ is preserved at finite temperature.
 Therefore for local field theory we expect that the partition function at finite temperature takes the form 
\begin{equation}\label{H3space} \log Z = \beta{\rm Vol}(\mathbb{H}_3)\mathcal F(\beta) \end{equation}
It remains to identify 
\begin{gather}
  {\rm Vol}(\mathbb{H}_3)= \int dv du  dy = \int \frac{\sin \theta \cos\theta d\theta d\phi_1 d\phi_2}{\gamma(\theta)^2} + \mathcal{O}(\epsilon_{1,2}) = \frac{\pi^2}{2\epsilon_1 \epsilon_2} + \mathcal{O}(\epsilon_{1,2})
\end{gather}
and the function $\mathcal F(\beta)$ appearing in~\eqref{H3space} is identified with the partition function in the large spin limit~\eqref{4duniversality}. Therefore, as in 2+1 dimensions, the extensive behavior of the large spin partition function is derived from locality of the quantum field theory on the space~\eqref{4dpp}.

\subsubsection{A Unitarity Bound in 3+1 Dimensions}

Let us first explain  that, if the Hamiltonian defined with respect to the metric
\eqref{eq:orig4dpp} is positive definite, one would conclude that
\begin{equation}
\tau = \Delta - J_1 - J_2 \geq  0 \, .
\end{equation}
For all operators in the CFT, and not just large spin operators. 
(This condition is stronger than the usual unitarity bounds, which read
$\Delta - \max(J_1,J_2) - 2 \ge 0$ for $J_1 \neq J_2$, or $\Delta - J - 1 \ge 0$
for $J_1 = J_2 = J$.)
To see this, suppose that the spectrum contains an operator with negative twist,
which we denote by $\mathcal O$.
Then one can construct an infinite family of operators with arbitrarily negative
twist by considering composite operators of the form
\begin{equation}
\mathcal O\, \partial^{\ell_1} \mathcal O\, \partial^{\ell_2} \mathcal O \cdots
\partial^{\ell_n} \mathcal O \, .
\end{equation}
As discussed previously, at sufficiently large spin the theory becomes
semiclassical, and the twist is approximately additive.
As a consequence, the existence of a single operator with negative twist would
imply the presence of operators with arbitrarily negative twist, rendering the
Hamiltonian defined on the metric \eqref{eq:orig4dpp} unbounded from below.

This is why the conjectured inequality $\tau \ge 0$~\cite{Cordova:2017dhq} follows from having a bounded from below energy in~\eqref{eq:orig4dpp}. 

Since the metric \eqref{eq:orig4dpp} admits a timelike Killing vector
$\partial_t$ everywhere in spacetime,
a classical Hamiltonian is positive definite.
There is no ergosphere. 
 One similarly expects a quantum system to have a bounded from below energy in this spacetime. We will now explain how this follows from causality.

Our discussion so far was appropriate for the thermodynamics with fixed $J_1/J_2$ with $J_1,J_2\to\infty$.
We could also discuss the problem of $J_1\to\infty$ with $J_1\gg J_2 $.
Then the partons are spinning fast only in the $\phi_1$ cycle and the derivation of the appropriate metric is essentially identical to the procedure in 2+1 dimensions since we are now forced to zoom in near $\theta=\pi/2$. In fact, when only one angular momentum large the discussion is pretty much identical across dimensions. So let us denote by $d+1$ the number of space-time dimensions then after manipulations similar to those in 2+1 dimensions we obtain the following metric
\begin{gather}\label{onelarge}
    ds^2 = -\left(1 + \sum^{d-1}_{i=1}x_i^2\right) dt^2 + dx dt + \sum^{d-1}_{i=1} dx_i^2~, \quad {\rm Vol}_d = \frac{\pi^\frac{d+1}{2}}{\epsilon \Gamma(1/2+d/2)}~.
\end{gather}
We conclude  that the free energy, in the case of a single large angular momentum, takes the form $\log Z \propto {\beta \over \epsilon}\mathcal F(\beta) $ in all space times dimensions above 1+1. 
The latter  metric has an isometry group given by a Heisenberg group \(\mathbb H_{2(d-1)+1}\) (acting on the transverse directions and the null coordinate $x$) together with time translations. The only generators that commute with $\partial_t$ are translations along the $x$ and $t$ directions, together with rotations of the transverse coordinates $x_i$. 

 As we show in appendix~\ref{app:ckv}, the Hamiltonian of a CFT placed in the metric \eqref{onelarge} could be expressed through the conformal generators in the following way
\begin{gather}
    H = P_u + P_v - K_v~.
\end{gather}
Using the algebra of CKVs, it is possible to show that $H\geq 0$.\footnote{The metric in Eq.~\eqref{onelarge}  admits a natural interpretation within the framework of non-relativistic conformal field theory (NR-CFT). It was argued in Ref.~\cite{Goldberger:2008vg} that \eqref{onelarge} realizes the Schr\"odinger symmetry group, and an AdS dual description was subsequently investigated in Ref.~\cite{Maldacena:2008wh}.} This is not surprising since the geometry~\eqref{onelarge} is obtained by a scaling limit with one large angular momentum and it is known that the corresponding twist is positive \footnote{Shifting the twist by $P_v$ does not spoil the positivity, as in the 2+1 dimensional case we discussed before. This can be seen both from the algebra of CKVs or by appealing to causality.}.

 Note that, starting from the metric \eqref{onelarge}, one obtains the metric
\eqref{eq:orig4dpp} by the globally defined coordinate change, that is just a rotation with a constant angular velocity in the $x_{1,2}$ plane
\[
x_1 = u \cos t + y \sin t,\qquad
x_2 = -u \sin t + y \cos t,\qquad
x = v + u y \, .
\]
It follows immediately that the isometry groups of
\eqref{eq:orig4dpp} and \eqref{onelarge} coincide. The stabilizers of time
translations are different in the two cases; indeed, time translations generated by \eqref{eq:orig4dpp} admit a
larger stabilizer.

 We are now ready to complete the argument that $\tau=\Delta-J_1-J_2\geq 0$. For that let us note that we have started from the spacetime~\eqref{onelarge} where the energy is non-negative and there is a global timelike Killing vector. Then we are switching to a rotating frame in the $(x_1,x_2)$ plane. In flat space this leads to $g_{00}$ that flips sign due to the rotation exceeding the speed of light. But in our case this rotation never exceeds the speed of light since the effective speed of light grows quadratically with the distance from the origin. Therefore, causality enforces that $\tau=\Delta-J_1-J_2\geq 0$. Equivalently, in the rotating frame, we do not have negative energies.
As we explained, this implies the said unitarity bound for all operators, and not just large spin operators.

\section{Examples}\label{examples}
In the following we consider explicit examples in which we demonstrate the connection between the thermal partition function in the large-spin limit and the geometries proposed above.

\subsection{Free scalar field in 2+1 d}
The single-particle excitation spectrum of a free scalar field theory in \(2+1\) dimensions is
\begin{gather}
    \Delta_l = l + \frac12~, 
\end{gather}
with $l$ being the  $SU(2)$ spin.
From this we compute the single-particle partition function for the Hamiltonian
\begin{gather}
    H = \Delta - (1-\epsilon) J ,
\end{gather}
and obtain the free energy using the plethystic exponent. We take the $\epsilon\ll 1$ limit. The single-particle partition function is
\begin{gather}
    Z_{\rm sp}
    = \sum_{l=0}^{\infty} e^{-\beta\left(l+\frac12\right)}
    \frac{\sinh \frac{\beta(1-\epsilon)(2l+1)}{2}}{\sinh \frac{\beta(1-\epsilon)}{2}}
    = \frac{1}{2\epsilon\,\beta\,\sinh \frac{\beta}{2}} \ .
\end{gather}
Therefore the partition function is
\begin{gather}\label{opcount}
     \log Z(\beta,\epsilon)= \sum_{m=1}^{\infty}\frac{1}{2\epsilon\,\beta\,m^2\,\sinh \frac{\beta m}{2}} \ .
\end{gather}
Let us now compute the spectrum of a free conformal scalar field in this space-time.  An early paper showing that scalar fields are well behaved in such spacetimes is~\cite{Gibbons:1975jb}. We write the equation of motion for the scalar field and determine the corresponding normal modes and their spectrum. The equation of motion is
\begin{gather}
    (1+y^2)\,\partial_x^2\phi + \frac14 \partial_y^2\phi + \,\partial_t\partial_x\phi = 0 .
\end{gather}
Substituting the plane-wave ansatz \(\phi = e^{ikx}\,\varphi(t,y)\) with \(k>0\), we find
\begin{gather}
    i\,\partial_t \varphi
    = -\frac{1}{4k}\,\partial_y^2 \varphi + \Bigl(1+y^2\Bigr)k\,\varphi ,
\end{gather}
where $k>0$.
This is a Schr\"odinger equation for a particle of mass \(2k\) in a harmonic oscillator potential with frequency \(\omega = 1\). Stability requires \(k>0\). The corresponding eigenmodes have energies
\begin{gather}
    E_{n,k} = k + \left(n+\frac{1}{2}\right) .
\end{gather}
From this we can easily compute the thermal free energy of our space.
We obtain for the single particle states in the pp-wave geometry 
\begin{equation}Z_1=V\int_{0}^\infty {dk\over 2\pi}\sum_n e^{-\beta(k+n+\frac12)}= {V\over 4\pi}{1\over \beta \sinh(\beta/2)} \end{equation} 
Now identifying the volume of the pp-wave with 
\(V = \frac{2\pi}{\epsilon}\), we obtain via the Plethystic exponential
\begin{gather}
  \log Z(\beta,\epsilon)= \frac{1}{\epsilon\,\beta}\sum_{m=1}^{\infty}\frac{1}{2m^2\,\sinh\frac{m\beta}{2}}~,
\end{gather}
which matches the result from direct  operator counting~\eqref{opcount}. The identification \(V = \frac{2\pi}{\epsilon}\) between the pp wave  volume and the operator counting problem is model independent. 

We can also study the case of \(H = \Delta - (1-\epsilon)J_{\star}\) in \(d\) dimensions, where $J_\star$ is the spin conjugate to one of the angles. (So we only take one angular momentum to be large, in the $\epsilon\ll1$ limit.)

The treatment of this problem is via the geometry~\eqref{onelarge}. The results are analogous to the three-dimensional case, but now, instead of one harmonic oscillator, we have \(d-2\) of them. To show this, let us write the equation of motion for a free conformal scalar in \(d\) dimensions,
\begin{gather}
    \left[\partial_t \partial_x + \left(1 + \sum^{d-2}_{i=1} x_i^2\right)\partial_x^2 + \frac14 \Delta_{d-2}\right] \phi = 0
\end{gather}
Using the ansatz \(\phi = e^{i k x} \prod\limits^{d-2}_{i=1} \psi_i(x_i)\), we see that this equation decouples and
\begin{gather}
    E = k +\sum^{d-2}_{i=1}\left(n_i + \frac12\right), \qquad
   \log Z(\beta,\epsilon) = \frac{1}{\epsilon \beta}\sum^\infty_{m=1}\frac{1}{m^2 \left(2 \sinh \frac{m \beta}{2}\right)^{d-2}}  \ .
\end{gather}
A quick consistency check that this indeed agrees with the direction calculation for a scalar field, we use the large temperature expansion to get
\begin{gather}
    \log Z(\beta,\epsilon) = \frac{\zeta(d)}{\beta^{d-1} \epsilon} = -\frac{\pi^\frac{d}{2} \beta}{\epsilon \Gamma(d/2)} F_{d}(\beta),
\end{gather}
where $F_{d}(\beta) = -\frac{\Gamma(d/2)\zeta(d)}{\pi^\frac{d}{2}\beta^d}$ is the free energy of a massless scalar field in $d$ dimensions in flat space-time.

\subsection{Free fermion in 2+1 d}
The spectrum of single particle excitations of a free Dirac fermion on $S^2 \times S^1$ is given by
\begin{gather}
    \Delta_l = l~,\quad  SU(2): \left(l+\frac12\right) \oplus \left(l+\frac12\right) \ .
\end{gather}
We can compute in the $\epsilon\ll1$ limit
\begin{gather}
    Z_{\rm sp} = 2\sum^\infty_{l = 1} e^{-\beta\left(l + \frac12\right)} \frac{\sinh\left(\beta(1-\epsilon) l\right)}{\sinh \frac{\beta(1-\epsilon)}{2}} = \frac{1}{\beta\epsilon \sinh \frac{\beta}{2}}
\end{gather}
And using the plethystic exponent for fermions we get
\begin{gather}
    \log Z(\beta,\epsilon) = \frac{1}{\epsilon}\sum_m  \frac{(-1)^m}{m^2\beta \sinh \frac{m\beta}{2}} \ .
\end{gather}
Now let us show that the same free energy could be obtained by considering a theory of a free Dirac fermion in the pp-wave background. For that we  need to study the spectrum of the Dirac operator in this background. 
We have that 
\begin{equation}
    \slashed{ D} \psi = \left(\gamma^t (\partial_t+(2 + 2 y^2)\partial_x)+\gamma^x \partial_t+ \frac{1}{\sqrt{2}}\gamma^y \partial_y\right)\psi = 0 ~.
\end{equation}
For simplicity we  square this operator to get
\begin{equation}
     \slashed{D} ^2 \psi =  \left(-2 \partial_t\partial_x-2(1+y^2)\partial_x^2+\frac12 \partial_y^2+\frac{\sqrt{2}y}{2} \gamma^+\gamma^2 \partial_x\right) \psi =0 \ ,
\end{equation}
note that $\gamma^+ = \frac{1}{\sqrt{2}}\left(\gamma^0 + \gamma^1\right)$ is nilpotent, and hence from the structure of this equation we expect that the solution should have the following form
\begin{equation}
    \psi = \gamma^+ \chi(y) e^{i \omega t+i k x} \ ,
\end{equation}
from which
\begin{equation}
   \left[ \omega k+(1  +y^2) k^2+ \frac14 \partial_y^2\right] \chi(y) = 0 \ .
\end{equation}
Therefore the spectrum of this operator is just the same as in the previous sub-section
 \begin{equation}
    E_{n,k} =  k+ n+ \frac12 \ ,
\end{equation}
positivity of energy imposes $k>0$. Using the plethystic exponent for the fermions we can get that the free energy in this case is 
\begin{gather}
   \log Z(\beta,\epsilon) %
    = \sum_{m=1}^\infty \frac{(-1)^m}{\epsilon \beta m^2 \sinh\frac{m \beta}{2}} \ .
\end{gather}
Once again the result perfectly matches the operator counting.

We see that the spectrum of the free scalar and free fermion coincide. As we mentioned in the introduction, this should be attributed to having supercharges commuting with the twist generator.

\subsection{Free scalar field in 3+1 d} 
Now we will study various free conformal theories in 3+1 d. And we will start with the free conformal scalar. The spectrum of single particle excitations in this case is
\begin{gather}
    \Delta_l = 1 + l, \quad SU(2)_L \times SU(2)_R : \left(\frac{l}{2},\frac{l}{2}\right),
\end{gather}
and from that we can find that the partition function with respect to the Hamiltonian $H = \Delta - 2J_R  + (\epsilon_1 + \epsilon_2) J_R + (\epsilon_1 - \epsilon_2)J_L$. Then we can find that the free energy to be, in the $\epsilon_{1,2}
\ll1$ limit
\begin{gather*}
    Z_{\rm sp} =\sum^\infty_{l=0} e^{-\beta(l+1)} \frac{\sinh\left[ \frac{1}{2}(l + 1) \beta(2 - \epsilon_1 - \epsilon_2) \right]}{\sinh \left[\frac12\beta\left(2 - \epsilon_1 - \epsilon_2\right) \right]}  \frac{\sinh\left[ \frac{1}{2} (l + 1) \beta( \epsilon_1 - \epsilon_2) \right]}{\sinh\left[\frac12 \beta(\epsilon_1 - \epsilon_2)\right]} = \frac{1}{2\epsilon_1 \epsilon_2 \beta^2\sinh  \beta } + \ldots,
\end{gather*}
And from that we find that the free energy is
\begin{gather}
    \log Z(\beta,\epsilon)= \frac{1}{2 \epsilon_1 \epsilon_2 \beta^2 } \sum_m \frac{1}{m^3 \sinh m\beta} 
\end{gather}
Let us try to match to the computation in the pp wave background. For that we just need to find a spectrum of free conformal scalar field in background defined in \eqref{4dpp} by just solving the equations of motions of a scalar field
\begin{equation}
\left(
-2\partial_t\partial_v
+(4u^2+1)\,\partial_v^2
-4u\,\partial_v\partial_y
+\partial_u^2
+\partial_y^2
\right)\phi=0 .
\end{equation}

Substituting the ansatz $
\phi = \varphi(u) e^{i p v + i k y + i \omega t'}$ we again get just a Schr\"odinger equation for a free particle in a external constant magnetic field
\begin{equation}
i\partial_{t'}\varphi=
\left[
\frac{p}{2}-\frac{\partial_u^2}{2p}
+2p\left(u-\frac{k}{2p}\right)^2
\right]\varphi .
\end{equation}
The spectrum is \begin{equation}
    E_n = \frac{p}{2}+2n+1 \ ,
\end{equation}
where $p>0$.
Let us emphasize that the energy eigenvalues are independent of $k$. This feature follows from the $\mathbb H_3$ symmetry: states with different values of $k$ are related by the action of $\mathbb H_3$, and therefore share the same energy. The label $k$ thus parametrizes the degeneracy associated with the Heisenberg group.

From the energy spectrum we construct the partition function 
in the same way as we did in the 2+1 dimensional case, to find
\begin{gather}
\log Z(\beta,\epsilon) = \frac{L_v L_y L_u}{\pi^2 \beta^2 } \sum_m \frac{1}{m^3 \sinh m\beta}~.
\end{gather}
Now using that $L_y L_uL_v = V = \frac{\pi^2}{2\epsilon_1 \epsilon_2}$. From that we find that
\begin{gather}
    \log Z(\beta,\epsilon)=  \frac{1}{2 \epsilon_1 \epsilon_2 \beta^2 } \sum_m \frac{1}{m^3 \sinh m\beta} \  .
\end{gather}

\subsection{Free Fermion in 3+1 d }
We now consider a free massless Dirac fermion in four dimensions. To simplify the computation in this case we will set $\epsilon_1 = \epsilon_2 = \epsilon$. The spectrum of single–particle excitations on $S^3\times S^1$ is well known. The eigenvalues of the Dirac Hamiltonian are
\begin{gather}
    \Delta_n = n+\frac{3}{2}, \qquad n=0,1,2,\dots
\end{gather}
and the corresponding $SU(2)_L\times SU(2)_R$ quantum numbers are
\begin{gather}
    SU(2)_L\times SU(2)_R:
    \left(\frac{n+1}{2},\frac{n}{2}\right)\oplus
    \left(\frac{n}{2},\frac{n+1}{2}\right).
\end{gather}
Using this spectrum, the single–particle partition function with a chemical potential $2\beta(1-\epsilon)$ for $J_L^z$ is
\begin{gather}
    Z_{\rm sp}(\beta,\epsilon)
    =
    \sum_{n=0}^{\infty} e^{-\beta\left(n+\frac{3}{2}\right)}
    \frac{(n+1)\sinh\!\left[(n+2)\beta(1-\epsilon)\right]
    +(n+2)\sinh\!\left[n\beta(1-\epsilon)\right]}
    {\sinh\!\left[\beta(1-\epsilon)\right]} .
\end{gather}
In the limit $\epsilon\to0$ this behaves as
\begin{gather}
    Z_{\rm sp}(\beta,\epsilon)
=\frac{1}{ 2 \epsilon^2 \beta^2  \sinh\frac{\beta}{2}}
    + O(\epsilon^{-1}) .
\end{gather}
The full fermionic free energy is then obtained by the standard plethystic exponent 
\begin{gather}
    \log Z(\beta,\epsilon)
    =
    \frac{1}{\epsilon^2}
    \sum_{m=1}^{\infty}
    \frac{(-1)^{m+1}}{ 2\beta^2  m^3\sinh\frac{m \beta}{2}}.
\end{gather}
The Dirac equation written explicitly in the coordinates
$(t,v,u,y)$ of the pp wave defined in \eqref{eq:orig4dpp} is
\begin{align}
\Bigg[
\gamma^+ \left(\partial_{t}- \frac12 \partial_v\right)
+ \gamma^- \partial_v + \gamma^1 \partial_u  + \gamma^2 \left(\partial_y - 2 u \partial_v\right) + \frac12 \gamma^{+12}\Bigg] \Psi = 0
\label{eq:ppwave-dirac-explicit}
\end{align}
Using the  ansatz
\begin{gather}
\psi(t,v,y,u)=e^{-i\omega t-i p v+i k y}\,\Psi(u)
\end{gather}
we arrive at the following equation
\begin{align}
\Bigg[
-i \gamma^+ \left(\omega - \frac{p}{2}\right)
- i \gamma^- p + \gamma^1 \partial_u  + i \gamma^2 \left(k + 2 u p \right) + \frac12 \gamma^{+12}\Bigg] \Psi = 0,
\label{eq:ppwave-dirac-explicit}
\end{align}
using that $\Psi = \Psi_+ + \Psi_-$ with $\gamma^+ \Psi_+ = \gamma^- \Psi_- = 0$ we get the following equation
\begin{gather}
    \Psi_+ = -\frac{(\gamma^1 \partial_u + i \gamma^2 (k + 2 u p))}{2 i p} \gamma^+\Psi_-
\end{gather}
Substiting it back into the above equation we find
\begin{gather}
    \omega \Psi_- = \frac{p}{2} \Psi_- + \left[ -\frac{\partial_u^2}{2p} + 2 p \left(u + \frac{k}{2p}\right)^2\right] \Psi_- + \frac{i}{2}\gamma^{12} \Psi_-\,,
\end{gather}
and from that we find that the spectrum is
\begin{gather}
    E = \frac{p}{2} + 2 n + 1 + \frac{s}{2}, \ .
\end{gather}
{Supercharges have $\Delta = 1/2$, $J_1 = 1/2$, $J_2 = \pm 1/2$ and therefore they do not commute with the Hamiltonian $\Delta-J_1-J_2$.}
where $s = \pm 1$ is the eigenvalue of $\gamma^{12}$.
As in the free scalar case, we observe that the energy eigenvalues do not depend on the parameter $k$. This is a direct manifestation of the $\mathbb H_3$ symmetry.
And now we can find that 
\begin{gather}
    Z_{\rm sp} = L_y L_v \sum_{n,s} \int \frac{dp}{2\pi}\, \int^{2p L_u}_0 \frac{dk}{2\pi} \, e^{-\beta p/2 - \beta (2 n + 1)- \beta s/2} = \frac{V}{\pi^2  \beta^2 \sinh \frac{\beta}{2}} \ .
\end{gather}
The free energy is
\begin{gather}
   \log Z(\beta,\epsilon) = \frac{1}{\epsilon^2}\sum_m \frac{(-1)^m}{2m^3 \beta^2 \sinh \frac{m \beta}{2}} ~,
\end{gather}
in agreement with the explicit operator counting.
\subsection{Free Maxwell theory in 3+1 d}
Another free theory that could be studied explicitly is a $U(1)$ gauge theory in $3+1$d. Again for simplicity we will consider that $\epsilon_1 = \epsilon_2 = \epsilon$. The spectrum of single particle excitations of this theory on $S^3 \times S^1$ is given by
\begin{gather}
    \Delta_l = l + 1, SU(2) \times SU(2): \left( \frac{l + 1}{2}, \frac{l-1}{2}\right) \oplus \left( \frac{l - 1}{2}, \frac{l+ 1}{2}\right) 
\end{gather}
From this we can find 
\begin{gather*}
    Z_{\rm sp}(\beta, \epsilon) = \sum_{l} e^{-\beta (l+1)} \frac{l \sinh \left[\left(l + 2\right)\beta(1-\epsilon)\right] + (l+2) \sinh \left[l \beta(1-\epsilon)\right]}{\sinh \beta(1-\epsilon)} = \frac{\coth \beta}{\beta^2\epsilon^2} + \ldots
\end{gather*}
And we can compute the following free-energy in this limit to be
\begin{gather}
    \log Z(\beta,\epsilon) =\frac{1}{\epsilon^2} \sum  \frac{\coth n\beta}{n^3 \beta^2} + \ldots
\end{gather}
Now using the metric of pp-wave we can compute the free energy of Maxwell field in that background. For that we write Maxwell's equations in the background \eqref{eq:orig4dpp}. In the lightcone gauge $A_v = 0$ and choosing the ansatz $A_\mu = a_\mu(y,t) e^{i p v + i k y}$ we get the following independent equations
\begin{gather}
p a_t + (2p u - k) a_y + i \partial_y a_u = 0, \notag\\
     -i\partial_t a_u =  \frac{p}{2} a_u- \frac{\partial_y^2}{2 p} a_u + 2p \left( u - \frac{k}{2p}\right)^2 a_u + i a_y, \notag\\
     -i\partial_t a_y = \frac{p}{2} a_y- \frac{\partial_y^2}{2 p} a_y + 2p \left( u - \frac{k}{2p}\right)^2 a_y - i a_u,
\end{gather}
Introducing $a_\pm = a_u \pm i a_y$ we get that the first equation determines $a_t$ through $a_y$ and $a_u$ while the second equation gives the following Schrodinger problem
\begin{gather}
    i\partial_t a_\pm = \frac{p}{2} a_\pm- \frac{\partial_y^2}{2 p} a_\pm + 2p \left( u - \frac{k}{2p}\right)^2 a_\pm \pm a_\pm
\end{gather}
Giving the spectrum
\begin{gather}
    E_{\pm,n,p} = \frac{p}{2} + \left(2n+1\right)\pm1, \quad Z_{\rm sp} = \sum_{n,s} L_yL_v \int\frac{dp}{2\pi} \int\limits^{2pL_u}_0 \frac{dk}{2\pi} e^{-\frac{\beta}{2}p - (2 n + 1 \pm s) \beta} =  \frac{2V \coth \beta }{\pi^2\epsilon^2 \beta^2} \notag\\
    \log Z = \frac{2V}{\pi^2\epsilon^2 } \sum_m \frac{\coth m\beta}{m^2 \beta^2}
\end{gather}
again taking into account that we have Landau levels we can compute the free energy
\begin{gather}
    \log Z(\beta,\epsilon) =   \frac{1}{\epsilon^2}\sum_{m}  \frac{\coth m\beta}{\beta^2 m^3}   \ .
\end{gather}
Once again, let us emphasize that—as in the free scalar and free fermion cases—the $k$-independence of the spectrum is a manifestation of the $\mathbb H_3$ symmetry.

\subsection{Wilson--Fisher fixed point in $d=4-\varepsilon$ dimensions}
In this section we will consider the example of an interacting theory, where we can compute the corresponding free energy for $\tau = \Delta - J_{12}$ as a function of temperature $\beta$. For that we will consider usual Wilson-Fisher theory~\cite{Wilson:1971dc} in $d=4-\varepsilon$ dimensions in the background of a pp wave geometry corresponding to fast rotation in one plane
\begin{gather}
    S=\int d^d x\,\sqrt{g}\left[\frac12 (\partial_\mu\phi)^2+\frac{\lambda}{4!}\phi^4\right]\ , 
    \qquad 
    ds^2=-\Bigl(1+\sum_{i=1}^{d-2} x_i^2\Bigr)dt^2+\,dx\,dt+\sum_{i=1}^{d-2} dx_i^2\ ,
\end{gather}
where we have also used that the Ricci scalar vanishes $R=0$. At leading order in $\varepsilon\equiv 4-d$, the contribution to the free energy comes from two sources: the dependence of the free theory on the dimension, and the first-order correction in perturbation theory in $\lambda$ (equivalently, in $\varepsilon$), since the fixed point is located at \cite{Wilson:1971dc,Henriksson:2022rnm}
\begin{equation}
    \frac{\lambda_\star}{(4 \pi)^2}=\frac{\varepsilon}{3}+\mathcal{O}(\varepsilon^2)\ .
\end{equation}
Using that 
\begin{gather}
-\log\operatorname{tr} e^{-\beta H_0 - \beta V} =  F_0  +\beta \langle V \rangle_{\beta,0} + \ldots,
\end{gather}
we see that the first correction in perturbation theory is
\begin{gather}
   \delta F  = \frac{\lambda \beta}{32} \int d^{d-1}x \langle{\phi^2\rangle}^2_T,
\end{gather}
where we used that in the leading order in $\varepsilon$ our theory would have correlators that could be computed using Wick's theorem.
\begin{gather}
    \langle{\phi^2\rangle}_T = \sum_{n_i} \int \frac{dp}{2\pi} \frac{1}{p}\frac{\prod^{d-2}_{i=1}\psi_{n_i}^2(x_i)}{e^{\beta \left(p + \sum\limits^{d-2}_{i=1} \left(n_i + \frac12\right)\right)} - 1} =
    \sum^\infty_{m=1} \int \frac{dp}{2\pi} \frac{e^{-\beta  m p}}{p} \prod\limits^{d-2}_{i=1} \left(\psi^2_{n_i}(x_i) e^{-m\beta(n_i+\frac12)}\right) = \notag\\
      \sum^\infty_{m=1} \int \frac{dp}{2\pi^2} \frac{e^{-p \beta  m  - 2 p r^2 \tanh\left(\frac{m\beta }{2}\right)}}{\sinh  m\beta}   =  \sum^\infty_{m=1} \frac{1}{2\pi^2\sinh  m\beta}  \frac{1}{m \beta + r^2 \tanh\frac{m \beta}{2}} \approx \frac{1}{12\,(1+r^2) \beta^2}
\end{gather}
Then the answer for the first correction to the partition function in the $\varepsilon$ expansion is
\begin{gather*}
\begin{aligned}
\log Z(\beta, \epsilon) &= \frac{1}{\epsilon} \sum_{m} \frac{1}{m^{2} \beta \left( 2 \sinh \frac{m \beta}{2} \right)^{d-2}} \\
&\quad + \frac{\varepsilon}{12 \epsilon} \sum_{m,n=1}^{\infty} \frac{1}{\sinh n\beta \sinh m \beta} \left[ \frac{\ln \left( \frac{n \tanh \frac{m\beta}{2}}{m \tanh \frac{n\beta}{2}} \right)}{n \tanh \frac{m\beta}{2} - m \tanh \frac{n\beta}{2}} \right]
\end{aligned}
\end{gather*}
Note that, as in the free-field examples, the function \(F(\beta)\) is analytic for any value of \(\beta\). Therefore, we conclude that at first order in \(\varepsilon\) there is no phase transition in the twist spectrum at large spin.

Let us see how this function behaves for small $\beta$
\begin{gather}
    \log Z = \frac{\zeta(4-\varepsilon)}{\beta^{3-\epsilon} \epsilon} + \frac{\pi^4 \varepsilon}{216 \beta^3 \epsilon} =  -\beta  F_{4-\varepsilon}(\beta) V_{4-\varepsilon},
\end{gather}
where $F_{4-\varepsilon}$ is the free energy of WF theory in $d=4-\varepsilon$ dimensions at inverse temperature $\beta$. And the low temperature behavior of this function yields
\begin{gather}
    \log Z = \frac{1}{2^{d-2}\beta \epsilon} e^{-\beta\frac{d-2}{2}} + \frac{\varepsilon}{3\epsilon} e^{-2\beta},
\end{gather}
we see that the leading asymptotics did not change, implying that $\tau_{\rm min}$ does not get corrections in the first order of $\epsilon$ expansion. This is as expected.

The same computation can be carried out for the \(\mathrm O(N)\) model in the \(\varepsilon\)-expansion by considering \(N\) scalar fields and evaluating the theory at the $\mathrm O(N)$ fixed point, where \cite{Henriksson:2022rnm}
\[
\frac{\lambda_\star}{(4\pi)^2}=\frac{3\varepsilon}{N+8}+\mathrm O(\varepsilon^2)\,.
\]
That yields
\begin{gather*}
\begin{aligned}
\log Z(\beta, \epsilon) &= \frac{N}{\epsilon} \sum_{m} \frac{1}{m^{2} \beta \left( 2 \sinh \frac{m \beta}{2} \right)^{d-2}} \\
&\quad + \frac{\varepsilon N(N+2)}{4 \epsilon (N+8) } \sum_{m,n=1}^{\infty} \frac{1}{\sinh n\beta \sinh m \beta} \left[ \frac{\ln \left( \frac{n \tanh \frac{m\beta}{2}}{m \tanh \frac{n\beta}{2}} \right)}{n \tanh \frac{m\beta}{2} - m \tanh \frac{n\beta}{2}} \right]
\end{aligned}
\end{gather*}
One again finds that \(F(\beta)\) remains analytic for any value of \(\beta\) and for any \(N\).

\acknowledgments
We thank Ibrahima Bah, Justin Kulp, Jeremy Mann, Jake McNamara, Shiraz Minwalla, Nikita Nekrasov, and Leonardo Rastelli for useful discussions. We thank especially Ofer Aharony, Gabriel Cuomo and  Shiraz Minwalla for important and useful comments on the draft.  Z.K. gratefully acknowledges
NSF Award Number 2310283.

\appendix

\section{Conformal Algebra and Symmetries of pp-waves}
\label{app:ckv}
Consider the Brinkmann pp-wave metric:
$$
ds^2=-\sum_{i=1}^{d-1}x_i^2\,dt^2+dt\,dx+\sum_{i=1}^{d-1}dx_i^2,
$$
To reveakl the conformal structure of this spacetime, we introduce the following coordinate transformations:
$$
u=\tan t,\qquad v=r^2\tan t -x,\qquad y_i=\frac{x_i}{\cos t}
$$
Under this coordinate transformation, the metric takes a manifestly conformally flat form:
\begin{equation}\label{eq:metriceqflat}
    ds^2=\cos^2 t\,(-\,du\,dv+dy_i^{\,2})
\end{equation}
Because the metric is conformally equivalent to the flat metric, its conformal Killing algebra is maximal and isomorphic to $\mathfrak{so}(d+1,2)$. We can construct an explicit basis for this algebra by pulling back the standard conformal generators of the flat metric. 
To express these generators in the original $(t,x,x_i)$ coordinates, we will use that 
\begin{align*}
\partial_u &=  \cos^2 t\,\partial_t-\sin t\cos t\,x_i\partial_{x_i}+r^2\cos 2t \,\partial_x \\
\partial_v &= -\partial_x\\
\partial_{y_i} &= \cos t\,\partial_{x_i}+2x_i\sin t\,\partial_x 
\end{align*}

Using these relations, the full $(d+2)(d+3)/2$-dimensional Conformal Killing Vector (CKV) basis---with $u,v,y_i$ understood as functions of $(t,x,x_i)$---can be written as follows:
\begin{itemize}
    \item[$\star$] the usual \textit{translations} take the form
\begin{gather}
P_u=-i\partial_u,\quad P_v = -i\partial_v,\quad P_i=-i\partial_{y_i} \ ,
\end{gather}
Because of our choice of coordinates we must have $-i P_{u,v} \geq 0$.
\item[$\star $] \textit{Rotations} and \textit{Boosts}
\begin{gather}
    M_{ij}=y_i\partial_{y_j}-y_j\partial_{y_i}=x_i\partial_{x_j}-x_j\partial_{x_i}, \notag\\
M_{uv}=\frac12 u\partial_u-\frac12v\partial_v,\qquad M_{ui}=-\frac12 v\partial_{y_i}-y_i\partial_u,\qquad M_{vi}= - \frac12 u\partial_{y_i}-y_i\partial_v \ ,
\end{gather}
\item[$\star$] \textit{Dilation}
\begin{gather}
D=u\partial_u+v\partial_v+y_i\partial_{y_i} \ ,
\end{gather}
\item[$\star$] and finally \textit{Special Conformal Transformations}:
    \begin{gather} K_u=iv\,D+i(-uv+y^2)\partial_u,\quad K_v=iu\,D+i(-uv+y^2)\partial_v \notag\\
    K_i=-2iy_i\,D+i(-uv+y^2)\partial_{y_i} \ .
    \end{gather}
\end{itemize}

These vector fields satisfy the standard $\mathfrak{so}(d+1,2)$ commutation relations. Let us note the time generator in this basis is
\begin{gather}
    H =  P_u -  K_v
\end{gather}
Now we can do a similarity transformation
\begin{gather}
    e^{i\frac{\pi}{4} (P_u + K_v)} H e^{-i\frac{\pi}{4} (P_u + K_v)} = \left( D + 2 M_{uv}\right) \geq 0 
\end{gather}
that is following from a standard unitary bound, this Hamiltonian has an infinite degeneracy. For instance,
\begin{gather}
    \forall m \in \mathbb{N} ,\quad \partial_{+}^m \partial_z^n \phi, \quad H = n + \frac12,
\end{gather}
leads to the infinite degeneracy. We can remove this degeneracy by changing the Hamiltonian or changing our metric. Thus, in the case of the metric \eqref{onelarge} we have to make a transformation $x\to x + t$ that enforces the Hamiltonian to be
\begin{gather}
    H =  P_u + P_v -  K_v,
\end{gather}
that is again positive since $P_v \geq 0$. The other way to see it is to write an overlap of two states $|\Psi_f\rangle = \int d^d x f(x) \mathcal{O}(x) |0\rangle $ and see that unitarity demands that particular combination to be positive \cite{Mack:1975je}. We can make the same transformation to get
\begin{gather}
    H = \Delta + 2 M_{uv} + P_v, 
\end{gather}
because $[P_v, P_u] = [P_v, K_v] = 0$. And we see that $H$ now does not have an infinite degeneracy in the momentum space. 
Now we can find a Heisenberg subalgebra in this group. For that we pick our mass operator (i.e. the central operator) to be $M = P_v$. Then a convenient choice for the Heisenberg group is
\begin{align}
    P_i =  & i\, (\cos t \partial_{x_i} + 2 x_i \sin t \partial_x), \notag\\
    X_i = M_{v i } = & i\, \left( \sin t \partial_{x_i} -2 x_i  \cos t \partial_x  \right)
\end{align}
And the satisfy standard commutation relations
\begin{gather}
    [P_i , X_j] = \delta_{ij} P_\sigma, \quad \alpha_i^\pm = P_i \pm i X_i = e^{\pm i t} \left(\partial_{x_i} \pm 2 i x_i \partial_x\right), \quad [H,\alpha_i^\pm] = \pm \alpha^\pm_i
\end{gather}

\bibliography{Draft.bib}
\bibliographystyle{JHEP}

\end{document}

%% file: packages.tex
\usepackage{./jheppub}
\usepackage[utf8]{inputenc}
\usepackage[english]{babel}
\usepackage[T1]{fontenc}
\usepackage[titletoc]{appendix}
\usepackage{titletoc}
\usepackage{graphicx}
\usepackage{subcaption} %
\usepackage{xcolor}
\usepackage{amsmath,amssymb,amsthm}
\usepackage{bm}
\usepackage{float}
\usepackage[format=plain,labelfont={bf,it},textfont=it]{caption}
\usepackage{multirow}
\usepackage{slashed} %
\usepackage{tabularx}
\usepackage{bbold}
\usepackage[shortlabels]{enumitem}
\usepackage{tikz}
\usetikzlibrary{positioning,arrows,calc}
\usetikzlibrary{arrows.meta}
\usetikzlibrary{decorations.pathmorphing}
\usetikzlibrary{decorations.markings}
\usetikzlibrary{shapes.geometric}
\usetikzlibrary{patterns}
\tikzset{
  snake it/.style={
    decorate, 
    decoration=snake,
    segment length=3
  }
}
\usepackage{xparse}
\usepackage{pdfpages} %
\usepackage{titlesec} %
\usepackage{fancyhdr} %
\usepackage{emptypage} %
\usepackage{dsfont} %